\documentclass[prd,superscriptaddress,amsmath,twocolumn]{revtex4-1}
\usepackage{graphicx}
\usepackage{float}
\usepackage{epstopdf,cancel}
\usepackage{epsf,latexsym,bbm,euscript}
\usepackage{amssymb,amsmath}
\usepackage{mathtools} 
\usepackage{times,graphics}
\usepackage{soul,xcolor}
\usepackage[normalem]{ulem}
\usepackage{url,hyperref}
\hypersetup{colorlinks,linkcolor={blue!55!black},citecolor={red!50!black},urlcolor={blue!45!black},breaklinks=true}

\DeclareMathOperator\arctanh{arctanh}
\newcommand{\defeq}{\vcentcolon=}

\def\6{{\langle}}
\def\9{{\rangle}}

\newcommand{\be}{\begin{equation}}
\newcommand{\ee}{\end{equation}}
\newcommand{\ba}{\begin{eqnarray}}
\newcommand{\ea}{\end{eqnarray}}

 \newcommand{\beqa}{\begin{eqnarray}}
\newcommand{\eeqa}{\end{eqnarray}}

\def\be{\begin{equation}}
\def\ee{\end{equation}}

\def\bali{\begin{align}}
\def\eni{{\end{align}}}

\def\1{{{\mathbbm 1}}}

\def\half{{\tfrac{1}{2}}}

\def\pad{{\partial}}

\def\sg{\textsl{g}}

\begin{document}

\title{Black hole evaporation and semiclassical thin shell collapse}

\author{Valentina Baccetti}
\affiliation{Department of Physics and Astronomy, Macquarie University, Sydney, New South Wales 2109, Australia}
\affiliation{School of Science, RMIT University, Melbourne, Victoria 3000, Australia}
\author{Sebastian Murk}
\author{Daniel R. Terno}
\affiliation{Department of Physics and Astronomy, Macquarie University, Sydney, New South Wales 2109, Australia}

\begin{abstract}
	In case of spherical symmetry, the assumptions of finite-time formation of a trapped region and regularity of its
 boundary --- the apparent horizon --- are sufficient to identify the form of the metric and energy-momentum tensor in its vicinity.
 By comparison with the known results for quasistatic evaporation of black holes, we complete the identification of their parameters. Consistency
 of the Einstein equations allows only two possible types of higher-order terms in the energy-momentum tensor. By using its local conservation, we
  provide a method of calculation of the higher-order terms, explicitly determining the leading-order regular corrections. Contraction of a spherically symmetric thin
  dust shell is the simplest model of gravitational collapse. Nevertheless, the inclusion of a collapse-triggered radiation in different extensions of this
   model  leads to apparent contradictions. Using our results, we resolve these contradictions and show how gravitational collapse may be
   completed in finite time according to a distant observer.
\end{abstract}
\maketitle

\section{Introduction} \label{sec:intro}

Black holes were originally conceived as spacetime domains from where no  information can escape. Mathematically, they are defined as a complement of the causal past of future null infinity \cite{fn:book,he:book,poisson},
 and their null boundaries are event horizons. General relativity (GR) and many modified theories of gravity predict the formation of black holes at the final stage of gravitational collapse.
 Event horizons are global {teleological} entities that are generically unobservable \cite{visser:14}. Instead, locally defined surfaces provide a more suitable conceptual and {analytical} framework. 
A trapped region --- a spacetime domain where both radial null geodesics have negative expansion --- forms inside the collapsing matter. Its suitably defined outer boundary
(i.e., the apparent horizon or another related surface) asymptotically approaches the event horizon \cite{fn:book,he:book,poisson,visser:14,krishnam:14,faraoni:b}.

Classical matter that satisfies energy conditions \cite{he:book,econ} crosses the event horizon and reaches the singularity in finite proper time $\tau$
that we associate with an observer who is comoving with the collapsing matter (Alice). However, according to a distant outside observer (Bob),
horizon formation takes an infinite amount of time. The trapped region and its associated surfaces are hidden behind the event horizon and thus
cannot be observed by him. According to Bob, the collapsing matter remains in a perpetual state of approach to the event horizon.

Quantum effects, such as the emission of Hawking radiation, make the {underlying} physics more involved. Three features of Hawking radiation are relevant: it is triggered by the collapse and does not require horizons for its emission \cite{liberati,vsk:07,pp:09};
its energy-momentum tensor $T_{\mu\nu}\vcentcolon=\6\hat T_{\mu\nu}\9$ violates the null energy condition (NEC), i.e.,
there is a null vector $k^\mu$ such that $T_{\mu\nu}k^\mu k^\nu<0$ \cite{he:book,econ,bmps:95}; and it results in evaporation
(complete or to a Planck-scale remnant) of black holes in a finite time $t_E$ of a distant observer \cite{bmps:95,coy:15}.

In conjunction with causality, these three properties pose the following dilemma: either horizons are formed in {a finite time $t_\mathrm{S}$ as measured by Bob ($t_S<t_E$)} or they are not formed at all. In the latter case, the observed black hole candidates are actually exotic compact objects \cite{cp:na17}.
The former case may include the existence of a transient ({albeit} possibly long-lived) trapped region without event horizon and singularity \cite{qua-mod,hay:06, saini-stojkovic:2014}.

 Any of these features are physically relevant only if they are formed in finite time according to Bob. Given that any definition of a black hole involves trapped regions,
 we reformulate the assumption that a black hole exists as the statement that a trapped region (starting from a single marginally trapped surface) is formed at some finite time $t_S$.

The simplest setting to investigate is a spherically symmetric collapse, where the apparent horizon is unambiguously defined for all foliations that respect this symmetry \cite{aphor}.
The analysis of Ref.\ \cite{bmmt} that we utilize below is based on the assumptions of finite $t_S$ and regularity of the apparent horizon.
It produces explicit expressions for the energy-momentum tensor and the metric in the vicinity of an expanding or contracting trapped region.
Formation of the apparent horizon at finite $t_S$ requires violation of the NEC in its neighborhood. The simplest expressions for the metric in that neighborhood in the case of
expansion or contraction of the apparent horizon are given by the outgoing Vaidya metric with increasing mass and the ingoing  Vaidya metric with decreasing mass, respectively.
The expression in standard Schwarzschild coordinates $(t,r)$ is valid in both cases, but it includes a function of time that is set by the choice of the time variable. It is not determined
 by the local properties of the solutions of the Einstein equations alone. In Sec.~\ref{general}, we will determine this function for the case of
a macroscopic black hole in an approximately steady evaporation state.

Regularity considerations allow one to derive a generic limiting form of the energy-momentum tensor \cite{bmmt} when the radial coordinate approaches the apparent horizon.
In Sec.~\ref{subsec:allowed_form}, we show that there are only two possible forms of the higher-order terms and derive the first-order terms for the regular corrections.

The simplest model of gravitational collapse is the contraction of a massive infinitesimally thin dust shell that separates a flat interior region from a curved exterior.
This so-called thin shell formalism models narrow transition regions between spacetime domains as hypersurfaces of discontinuity.
Mathematical consistency is maintained by imposing junction conditions, i.e., rules for joining the solutions of the Einstein equations on both sides
of the hypersurface $\Sigma$ \cite{poisson}. For the collapse of a thin shell in an asymptotically flat spacetime, the interior geometry is flat,
and in classical GR, the exterior geometry is described by the Schwarzschild metric. The simplicity of the model allows one to obtain the explicit time dependence of the shell's radius $R(\tau)$ and to determine the point in time when the shell crosses the Schwarzschild radius, $R(\tau_c)=r_\sg=2M$. We briefly review the classical thin shell formalism in Sec.~\ref{shell-cl}.
   Thin shells are used to analyze various alternatives to black holes as the final stage of gravitational collapse \cite{vw:04,hay:06,bh-nohor,nyh:18}.
    This allows one to circumvent some of the controversial issues, such as the structure of the energy-momentum tensor within the collapsing body.
    Moreover, the exterior metric naturally has a Schwarzschild radius that is initially located within the Minkowski interior. Thus, one
     bypasses the problem of emergence of trapped surfaces and is able to model the emission of radiation that precedes their formation.

Nevertheless, the results that have been obtained so far appear contradictory. On the one hand, models that use the outgoing Vaidya metric, as well as general metrics that satisfy certain
regularity conditions, exhibit horizon avoidance \cite{kmy:13,ho:16a,bmt-1,cuwy:18,mnt}. That is, for an arbitrary law that
 describes conversion of the shell's mass to radiation, the gap $X \vcentcolon= R - r_\sg$ between the shell and the Schwarzschild radius remains positive at all times.
  However, the dynamics of these shells involves some peculiarities \cite{cuwy:18,mnt} that will be described below (Sec.~\ref{shell-plus}).
On the other hand, arguments that are based on the iterative evaluation of the effects of backreaction (starting with the results for the
energy-momentum tensor of Hawking radiation on the background of an eternal black hole) have shown that while the collapse duration is extended,
the shell eventually crosses the event horizon in a finite time of both Alice and Bob \cite{pp:09}.

Since the near-horizon geometry of an evaporating black hole is described by the ingoing Vaidya metric with decreasing mass,
this is the metric that should be used in thin shell models that aim to represent the last stages of gravitational collapse in the presence
of collapse-triggered radiation. This scenario is analyzed in Sec.~\ref{nege}. The implications of our findings are discussed in Sec.~\ref{summary}.

To simplify the notation, we label quantities on the shell $\Sigma$ by capital letters, e.g.\ $R \vcentcolon= r_{|\Sigma}$,  $F \vcentcolon= f(U,R)$. The jump of a physical quantity $A$ across the shell is
 written as $[A] \vcentcolon= A|_{\Sigma_+}-A|_{\Sigma_-}$. All derivatives are explicitly indicated by subscripts, as in $A_R=\pad_R A(U,R)$. The total proper time derivative $dA/d\tau$ is denoted
 as $\dot A$, and the total derivative over some parameter $\lambda$ is $A_\lambda \vcentcolon= A_R R_\lambda+A_U U_\lambda$.    The time $t$ always refers to the coordinate time (proper time of Bob at spacelike infinity).

\section{General properties of the metric near the Schwarzschild sphere}     \label{general}

We work within the framework of semiclassical gravity \cite{pp:09,bmt-1}. That means we use the concepts of GR, and quantum effects are taken into account via the semiclassical Einstein equations,
\be
R_{\mu\nu}-\half R g_{\mu\nu}=8\pi \6\hat{T}_{\mu\nu}\9, \label{semiein}
\ee
where $R_{\mu\nu}$ is the Ricci tensor and $\6\hat{T}_{\mu\nu}\9\equiv T_{\mu\nu}$ is the expectation value of the energy-momentum tensor.
The latter represents the entire matter content of the model; both the
 collapsing matter and the created quantum field excitations are included. This cumulative representation allows a self-consistent study of the dynamics without having recourse
  to   iterative calculations of the backreaction \cite{qua-mod}.

Three coordinate systems are particularly useful. We use the Schwarzschild radial coordinate $r$ and either the Schwarzschild time $t$ or
the retarded and advanced null coordinates $u$ and $v$, respectively. The most general spherically symmetric metric is given by
\begin{align}
 ds^2&=-e^{2h(t,r)}f(t,r)dt^2+f(t,r)^{-1}dr^2+r^2d\Omega, \label{sgenm}\\
 &=-e^{2h^u(u ,r)}f^u(u,r)du^2- 2e^{h^u(u,r)}du dr+r^2d\Omega, \label{efret}\\
 &=-e^{2h^v(v ,r)}f^v(v,r)dv^2+ 2e^{h^v(v,r)}dv dr+r^2d\Omega. \label{efad}
\end{align}
The function $f$ is coordinate independent, i.e., $f(t,r)=f^u(u(t,r),r)$, etc. \cite{ms,bardeen:81}, and we can decompose it as
\be
f=1-2M(t,r)/r=1-2M^u(u,r)/r.
\ee
Here, $M=C/2$ is the Misner-Sharp mass \cite{ms,faraoni:b}. It is invariantly defined via
\be
1-C/r\defeq\pad_\mu r\pad^\mu r.
\ee
We drop the superscripts on $f$ and $M$ in what follows as it does not lead to confusion. The functions $h$, $h^u$, and $h^v$ play the role of
integrating factors that turn, e.g., the expression
\be
dt=e^{-h}(e^{h^v}dv- f^{-1}dr) \label{intf}
\ee
into an exact differential, provided that the coordinate transformation exists \cite{bmmt}.

In an asymptotically flat spacetime, the time variable $t$ is the proper time of a stationary Bob; thus,
\be
\lim_{r\to\infty}h(t,r)=0, \quad \lim_{r\to\infty}f(t,r)=1.
\ee
In the following, we work in this setting, but our results are also applicable on a cosmological background if there exists an intermediate scale $r_\sg\ll r\ll L$, where $L$ is set by the cosmological model.

Two physically motivated assumptions result in the classification of the energy-momentum tensor and the resulting metrics. First, we assume that trapped regions
form at a finite time $t$ of Bob. This entails that the equation $f(z,r)=0$ has a solution. This solution, or, if there are several, the largest one, is the
  {Schwarzschild horizon radius} $r_\sg(z)$. Second, we require that the hypersurface $r=r_\sg$ is regular by demanding that the two curvature scalars
  that are obtained directly from the energy-momentum tensor,
\be
\mathrm{T} \defeq T^\mu_{~\mu}, \qquad \mathfrak{T} \defeq T^{\mu\nu}T_{\mu\nu},
\ee
are finite.

The Einstein equations that determine the functions $h$ and $C$  are
\begin{align}
G_{tt}=&\frac{e^{2h}(r-C)\pad_r C}{r^3}=8\pi T_{tt}, \label{gtt}\\
G_t^{\,r}=&\frac{\pad_t C}{r^2}=8\pi T_t^{\,r}, \label{gtr}\\
G^{rr}=&\frac{(r-C)(-\pad_r C+2(r-C)\pad_r h)}{r^3}=8\pi T^{rr}. \label{grr}
\end{align}

This is the simplest form of the equations. It provides a natural choice of the independent components of the energy-momentum tensor.
The metric of Eq.~\eqref{sgenm} entails  $T^\theta_{~\theta}\equiv T^\phi_{~\phi}$. Then, the trace and the square scalars of the energy-momentum tensor are
 \begin{align}
    \mathrm{T}=&-e^{-2h }T_{tt}/f  +T^{rr}/f +2T^\theta_{~\theta}, \label{fin1}\\
  \mathfrak{T}=&-2\left(\frac{e^{-h} T_t^{\,r}}{f }\right)^2+\left(\frac{e^{-2h }T_{tt}}{f }\right)^2
    +\left(\frac{T^{rr}}{f }\right)^2 +2\big(T^\theta_{~\theta}\big)^2.       \label{fin2}
    \end{align}
For future convenience, we introduce $\tau_t \defeq e^{-2h }T_{tt}$, $\tau^r \defeq T^{rr}$, and $\tau_t^r \defeq e^{-h} T_t^{\,r}$.

There are several possibilities for the energy-momentum tensor to satisfy these requirements. The generic one (that is consistent with
the known results of the energy-momentum tensor of Hawking radiation \cite{bmps:95,leviori:16,elster:83}) results in the limiting form of the $(tr)$ block of $T_{\mu\nu}$ \cite{bmmt},
\be
 T_{\hat{a}\hat{b}}=  \frac{\Xi}{f}   \begin{pmatrix}
1 & s  \vspace{1mm}\\
s   & 1 \end{pmatrix},
  \label {tneg}
\ee
where $s=\pm 1$ and we have used the orthonormal frame to simplify the expression. The Einstein equations have solutions that contain
a finite-time apparent horizon only if $\Xi=-\Upsilon^2\leqslant 0$. In the generic case $\Upsilon^2>0$, the leading terms in the metric functions are given by
\be
C=r_\sg(t)-a(t)\sqrt{x}+\frac{1}{3}x\ldots \label{c0sin}
\ee
and
\be
h=-\ln\frac{\sqrt{x}}{\xi_0(t)} +\frac{4}{3a}\sqrt{x}\ldots ,  \label{h2}
\ee
where $x\defeq r-r_\sg$ and $a\defeq 4\sqrt{\pi r_\sg^3}\,\Upsilon$. Here and in a similar setting below, the function $\xi_0(t)$ is not
determined by the equations but is set by the choice of the time variable \cite{bmmt}. In an asymptotically flat spacetime, it can be defined such that $h(t,r)\to 0$ at spacelike infinity.
For both expanding and contracting trapped regions, the comparison via Eq.~\eqref{gtr} of the divergent terms of  $\pad_t C$ and $8\pi e^h \tau ^t_r r^2$ allows one to identify
 \be
 {r_\sg}^\prime/{\xi_0}=\pm4\sqrt{\pi r_\sg}\,\Upsilon= \pm a/r_\sg,       \label{lumin}
 \ee
where the upper (lower) signs correspond to the growth (contraction) of the trapped region.

The energy-momentum tensor violates the NEC;
$T_{\hat{a}\hat{b}}k^{\hat a}k^{\hat b }<0$
for a radial null vector $k^{\hat a}=(1,s,0,0)$. The two possibilities --- growth and contraction of the trapped
 region --- are determined by the sign of $T_t^{~r}$. An evaporating black hole corresponds to the ingoing Vaidya metric
 with decreasing mass; i.e., using the advanced null coordinate $v$ the metric of Eq.~\eqref{efad} identifies
\be
	2h^v(v ,r)=0, \qquad C(v,r)=r_\sg(v), \qquad r_\sg'(v)<0.
\ee
The effective mass of a black hole can be defined as $M\defeq r_\sg/2$ \cite{bardeen:81}.
\section{Explicit form of the metric and energy-momentum tensor}
\subsection{Metric in the quasistationary case}\label{met-sta}

For the pure ingoing Vaidya metric with decreasing mass, the only nontrivial Einstein equation at the apparent horizon reads \cite{bmmt}
\be
\frac{dr_\sg(v)}{dv}=-8\pi \Upsilon^2 r_\sg^2.
\ee
On the other hand, for a macroscopic black hole ($r_\sg\gg 1$), the evaporation process is quasistationary and the results obtained on the background of an eternal black hole
 with the corresponding mass are expected to give a good approximation for many quantities, including the luminosity \cite{bmps:95,bardeen:81}. Hence,
 \be
 \frac{dM}{dv}\propto-M^{-2},
 \ee
 and thus
\be
\frac{dr_\sg(v)}{dv}\approx -\kappa/r_\sg^2, \label{page-v}
\ee
where $\kappa\sim 10^{-3}-10^{-4}$ \cite{bmps:95,elster:83,leviori:16}.  As a result,
\be
\Upsilon\approx\frac{\sqrt{\kappa}}{2 r_\sg \sqrt{2\pi}},
\ee
and
\be
\xi_0 \approx \sqrt{\frac{\kappa}{2 r_\sg}} \approx2\sqrt{\pi r_\sg^3}\Upsilon = \frac{a}{2}.
\ee

\subsection{Allowed forms of the energy-momentum tensor} \label{subsec:allowed_form}
 We now obtain higher-order terms in the series solution for $C(t,r)$ and $h(t,r)$, thereby extending the results of \cite{bmmt}.
Higher-order contributions to the components of the energy-momentum tensor can be given either as regular functions, e.g., for evaporation
 \begin{align}
 \tau_t&=-\Upsilon^2(t)+\sum_{n\geqslant 1}\alpha_n x^n, \label{regt}\\
 \tau^r_t&=-\Upsilon^2(t)+\sum_{n\geqslant 1}\beta_n x^n, \\
 \tau^r&=-\Upsilon^2(t)+\sum_{n\geqslant 1}\gamma_n x^n, \label{regr}
 \end{align}
 or as  regular-singular functions,
 \begin{align}
 \tau_t&=-\Upsilon^2(t)+x^k\sum_{n\geqslant 0}\alpha_n x^n,\\
 \tau^r_t&=-\Upsilon^2(t)+x^k\sum_{n\geqslant 0}\beta_n x^n, \\
 \tau^r&=-\Upsilon^2(t)+x^k\sum_{n\geqslant 0}\gamma_n x^n,
 \end{align}
for some $0<k<1$.

First, we will show that apart from the regular expansion only the case of $k=\half$ is consistent with the Einstein equations.
 A direct calculation shows that a series solution of Eq.~\eqref{gtt} exists for any $k$.
Set
\be
C=r_\sg-a\sqrt{x}\left(1-\frac{g_1}{a}x^k-\frac{g_2}{a}x^{2k}+\right)\ldots,
\ee
and Eq.~\eqref{gtt} becomes
\begin{align}
&-\frac{a}{2\sqrt{x}}+(k+\half)g_1x^{k-1/2}+\ldots \nonumber \\
&=-\frac{a}{2\sqrt{x}}+8\pi r_\sg^3\left(\frac{\alpha_0}{a}-\frac{\Upsilon^2g_1}{a^2}\right)x^{k-1/2}+\ldots,
\end{align}
and with matching powers of $x$ on both sides, it is possible to find a solution.

Equation \eqref{grr} for $h$ then becomes
\be
\pad_x h=-\frac{1}{2x}+8\pi r_\sg^3\left(\frac{\gamma_0}{a}-\frac{2\Upsilon^2g_1}{a^2}\right)x^{k-1}+\ldots,
\ee
allowing a series solution of the form
\be
	h = -\ln\frac{\sqrt{x}}{\xi_0(t)} + h_k x^k+\ldots .
\ee

Equation \eqref{gtr} describes the mass change and serves as a consistency check for the solutions $h$ and $C$, as
\begin{align}
r_\sg'-a'\sqrt{x}+\frac{ar_\sg'}{2\sqrt{x}}+g_1'x^{k+1/2}+(k-\half)r'_\sg g_1x^{k-1/2}+\ldots \nonumber \\
= 8\pi r_\sg^2(-\Upsilon^2+\beta x^k)\frac{e^{h_0}}{\sqrt{x}}\big(1+h_k x^k)+\ldots
\end{align}
has to hold, while it is possible for $k<1$ only if
\be
\half=2k-\half ;
\ee
hence, the only regular-singular case corresponds to $k=\half$.

Since $f\propto \sqrt{x}$ for $x\to 0$, for $k=\half$, the invariant $\mathrm{T}$ does not lead to any restrictions, but $\mathfrak{T}$ being finite requires
\be
2\beta_0=\alpha_0+\gamma_0 \label{const1}.
\ee

\subsection{Metric functions}
In both cases we have to consider the expansion follows the same pattern.
For a regular correction to $T_{\mu\nu}$, the Misner-Sharp mass has the form
\be
C(t,r_\sg+x)=r_\sg-a\sqrt{x}+\tfrac{1}{3}x+c x^{3/2}+g x^2+\ldots ,
\ee
where
\begin{align}
a&=4 \sqrt{\pi } r_\sg^{3/2}\Upsilon, \\
 c&= \frac{ \left(36 \pi
\alpha_1 r_\sg^3-108 \pi  r_\sg^2 \Upsilon^2-1\right)}{36 \sqrt{\pi } r_\sg^{3/2}
\Upsilon},\\
 g&=\frac{1}{540} \left(-\frac{36 \alpha_1}{\Upsilon^2}+\frac{1}{\pi  r_\sg^3
\Upsilon^2}+\frac{108}{r_\sg}\right) ,
\end{align} and
\be
h(t,r_\sg+x)=-\ln\frac{\sqrt{x}}{\xi_0}+k_2 \sqrt{x}+k_3 x+k_4 x^{3/2}+\ldots,
\ee where
\begin{align}
k_2 &= \frac{4}{3 a},\\
k_3 &= - \frac{3}{2r_\sg}-\frac{c}{a}+\frac{24 \pi  \alpha_1 r_\sg^3+24 \pi  \gamma_1 r_\sg^3-4  }{6 a^2},\\
k_4 &= \frac{2 \left( 27 a^2 g - 54 a c - 16 \right)}{81a^3} \nonumber \\
& \quad + \frac{2 \left( -54 a^2 + 144 \pi \alpha_1 r_\sg^4 + 144 \pi \gamma_1 r_\sg^4 \right)}{81a^3 r_\sg},
\end{align}
and the function $\xi_0(t)$ is determined by the limiting form of  $h$ for $x\to 0$ and in general cannot be recovered from the series solution.

For a correction that scales as $x^{1/2}$, we have at leading order
\begin{align}
	\tau_t&=-\Upsilon^2(t)+\alpha_0\sqrt{x}, \\
	\tau^r_t&=-\Upsilon^2(t)+\half(\alpha_0+\gamma_0)\sqrt{x}, \\
 	\tau^r&=-\Upsilon^2(t)+\gamma_0\sqrt{x}.
\end{align}

The metric functions in this case have the same structure. The leading-order corrections determine
\be
 C(t,r_\sg+x)=r_\sg-a\sqrt{x}+\bar b x+\bar c  x^{3/2}+\ldots
\ee and
\be
h(t,r_\sg+x)=-\ln\frac{\sqrt{x}}{\xi_0}+\bar k_2 \sqrt{x}+\bar k_3 x +\ldots,
\ee
where
\begin{align}
	\bar b&=\frac{1}{3}+  \frac{4 \sqrt{\pi }\alpha_0 r_\sg^{3/2}}{3 \Upsilon},\\
	\bar c&=\frac{8 \pi  \alpha_0^2 r_\sg^3-2 \sqrt{\pi }\alpha_0 r_\sg^{3/2}\Upsilon-108 \pi  r_\sg^2\Upsilon^4-\Upsilon^2}{36 \sqrt{\pi } r_\sg^{3/2}\Upsilon^3},
\end{align}
and
\begin{align}
	\bar k_2&=\frac{4}{3 a}-\frac{8 \pi  r_\sg^3 (\alpha_0-3 \gamma_0)}{3 a^2},\\
	\bar k_3&=-\frac{128 \pi ^2 \alpha_0 r_\sg^6 (\alpha_0-3 \gamma_0)}{9 a^4}+\frac{8 \pi  r_\sg^3 (7 \alpha_0-6 \gamma_0)}{9 a^3} \nonumber \\
	&\quad \hspace{0.5mm} -\frac{5}{9 a^2}-\frac{3}{4 r_\sg}.
\end{align}

\subsection{Higher-order terms in the expansion of the energy-momentum tensor}
The consistency requirement of Eq.~\eqref{gtr} and the conservation law
\be
\nabla_\mu T^{\mu\nu}=0
\ee
for $\nu=t$ and $r$ can be imposed order by order in powers of $x$. This allows one to identify higher-order terms in the expansion of $T_{\mu\nu}$, which can be expressed in terms of $\Upsilon$, $\xi_0$, and their derivatives.

Using the identity
\be
\nabla_\nu A_\mu^{\,\nu}=\frac{1}{\varsigma}\pad_\nu(\varsigma A_\mu^{~\nu})-\half \sg_{\alpha\beta,\mu}A^{\alpha\beta},
\ee
where $\varsigma=\sqrt{-\det{\sg}}$, which is valid for an arbitrary symmetric tensor, and the Einstein equations in the spherically symmetric case,
we get a simple form of the flux equation
\be
\nabla_\nu T_t^{~\nu}=\pad_t T^t_{~t}+r^{-2}\pad_r(r^2 T^r_{~t})=0. \label{consflux}
\ee
The radial component of the conservation law reads
\begin{align}
\nabla_\nu T_r^{~\nu}=
-\frac{1}{e^{h}}\pad_t\left(\frac{\tau^r_t}{f^2}\right) +\frac{1}{e^h r^2} \pad_r\left(\frac{e^h r^2\tau^r}{f}\right) \nonumber \\
+\pad_r\big(e^{2h} f \big) \frac{e^{-2 h}\tau_t}{2f^2} +\pad_r f\frac{\tau^r}{2 f^2}=0. \label{conr}
\end{align}
Below, we apply it to the regular case \big[Eqs.~\eqref{regt}--\eqref{regr}\big] and derive three independent equations for $\alpha_1$, $\beta_1$, and $\gamma_1$.

We first consider the terms beyond $1/\sqrt{x}$ in Eq.~\eqref{gtr}.  The coefficients of $x^0$ give no new equations, but the terms of $\sqrt{x}$ and $x$ give   
\be
\sqrt{\pi}\xi_0\big(r_\sg(3\alpha  -4\beta_1+\gamma_1)+8\Upsilon^2\big)-2\sqrt{r_\sg}\,\Upsilon'=0, \ee
and   \be
-2+3\pi r_\sg^2\big(r_\sg(19\alpha_1-60\beta_1+5\gamma_1)+48\Upsilon^2\big)=0,
\ee
respectively.

The flux equation does not lead to any more independent constraints on the leading-order corrections. The coefficients of the
inverse powers of $x$ in  Eq.~\eqref{conr} become identically zero if Eq.~\eqref{lumin} is satisfied, but terms with $x^0$ lead to
\be
\xi_0\big(r_\sg(11\alpha_1-12\beta_1-11\gamma_1)+48\Upsilon^2\big)-6\sqrt{r_\sg/\pi}\Upsilon'=0.
\ee

As a result,
\begin{align}
\alpha_1&=-\frac{7}{288\pi r_\sg^3}-\frac{13\Upsilon^2}{4 r_\sg} +\frac{35 \Upsilon'}{32\sqrt{\pi r_\sg}\,\xi_0}, \\
\beta_1 &=-\frac{11}{576\pi r_\sg^3}-\frac{\Upsilon^2}{8 r_\sg}+\frac{23\Upsilon'}{64\sqrt{\pi r_\sg}\,\xi_0}, \\
\gamma_1 &=-\frac{1}{288\pi r_\sg^3}+\frac{5\Upsilon^2}{4 r_\sg} +\frac{5 \Upsilon'}{32\sqrt{\pi r_\sg}\,\xi_0} .
\end{align}
In the quasistationary case, the dominant contribution comes from the first term on the right-hand side, as
\be
\frac{\Upsilon^2}{r_\sg} \sim \frac{  \Upsilon'}{\sqrt{\pi r_\sg}\,\xi_0}\sim\frac{\kappa}{r_\sg^5},
\ee
where we have used the results of Sec.~\ref{met-sta}.

\section{Thin shell dynamics} \label{shell}
We begin by  briefly reviewing the collapse of a classical  massive spherically symmetric thin dust shell $\Sigma$ in 3+1 dimensions using the thin shell formalism \cite{poisson,bmt-1}.  The
spacetime inside the shell is assumed to be flat, and the exterior geometry is described by the Schwarzschild metric.
Next, we consider two models that incorporate mass loss by the shell, modeling the  exterior geometry either by an outgoing Vaidya metric or an ingoing Vaidya metric with decreasing mass.
\subsection{Classical thin shell formalism}\label{shell-cl}
The metric across the two domains that the shell separates can be represented as the distributional tensor \be
\bar{\sg}_{\mu\nu}=\bar{\sg}_{\mu\nu}^+{\Theta}(\xi)+\bar{\sg}_{\mu\nu}^-\Theta(-\xi),
\ee
using the set of special coordinates $\bar x^\mu=(w,\xi,\theta,\phi)$. Here, $\Theta(\xi)$ is the step function, and the interior and
exterior metrics $\bar \sg^\pm(\bar x)$ are continuously joined at $\xi=0$. The coordinates $w$ and $\xi$ as well as the explicit
form of this metric in a general spherically symmetric case are given in Ref.\ \cite{mnt}.

A mathematically equivalent approach is the thin shell formalism. It is particularly convenient for our purposes. Birkhoff's theorem
imposes the Schwarzschild metric at the shell's exterior,
\begin{align}
ds^2_+ & =  -f(r_+) du^2_+ - 2du_+ dr_+ + r^2_+ d\Omega\nonumber \\
& = - f(r_+) dv^2_+ + 2dv_+dr_+ + r^2_+ d\Omega\nonumber \\
& =- f(r_+) dt^2_+ - f^{-1}(r_+) dr^2_+ + r^2_+ d\Omega ,
\end{align}
where the subscript $+$ denotes the exterior region and the retarded and advanced null coordinates $u$ and $v$ are the Eddington-Finkelstein (EF) coordinates.

The interior region is described by the Minkowski metric,
\begin{align}
ds^2 _-& = -du^2_- - 2 du_- dr_- + r^2_- d\Omega \nonumber \\
            & = -dv^2_- + 2 dv_- dr_- + r^2_- d\Omega \nonumber \\
            & = -dt^2_- + dr^2_- + r^2_- d\Omega,
\end{align}
where $u_-=t_--r_-$, $v_-=t_-+r_-$. The shell's trajectory is parametrized by the proper time $\tau$ as $\big(T_\pm(\tau), R_\pm(\tau)\big)$ or $\big(V_\pm(\tau), R_\pm(\tau)\big)$
using, respectively, $(t,r)$ or $(v,r)$ coordinates outside and inside the shell. We use the hypersurface coordinates $y^a=(\tau,\Theta \vcentcolon= \theta|_{\Sigma},\Phi \vcentcolon= \phi|_{\Sigma})$.
The first junction condition \cite{poisson}, which is the statement that the induced metric $h_{ab}$ is the same on the both sides of the shell $\Sigma$,
\begin{align}
	 ds^2_\Sigma=h_{ab}dy^ady^b=-d\tau^2+R^2d\Omega_{D-1} ,
\end{align}
leads to the identification $R_+\equiv R_- =\vcentcolon R(\tau)$. Henceforth, we drop the subscripts from the radial coordinate. Trajectories of the shell's particles are timelike; hence,
\begin{align}
\dot U_+ &= \frac{-\dot R +\sqrt{F+\dot R^2}}{F}, \label{uder} \\
\dot V_+ &= \frac{\dot R +\sqrt{F+\dot R^2}}{F}, \label{vder}\\
\dot T_+ &=\frac{\sqrt{F+\dot R^2}}{F},   \label{timeder}
\end{align}
where $F = 1-r_\sg/R$.  These expressions are applicable also for a general $C(z,r)$ (while $h\equiv 0$). In the following, we drop the subscript $+$ from the exterior quantities.

The surface energy-momentum tensor of a massive thin dust shell is
\be
S^{ab}=\sigma v^a v^b=\sigma \delta^a_\tau\delta^b_\tau,
\ee
where $\sigma$ denotes the surface density. The rest mass of the shell is $m=4\pi \sigma R^2$. The second junction condition relates the jump in extrinsic curvature
\be
K_{ab} \vcentcolon= \hat n_{\mu;\nu}e^\mu_a e^\nu_b \label{extK}
\ee
to the surface energy-momentum tensor
\be
S_{ab}=-\big([K_{ab}]-[K]h_{ab}\big)/8\pi, \label{eqofmot}
\ee
where $K \vcentcolon= K^a_{\,a}$ and $[K] \vcentcolon= K|_{\Sigma^+}-K|_{\Sigma^-}$ is the discontinuity of the extrinsic curvature scalar $K$ across the two sides $\Sigma^\pm$ of the surface.

The equation of motion for the shell can be obtained from
\begin{eqnarray}
  {\mathcal{D}(R) } & \vcentcolon= &  \frac{2\ddot R + F'}{2\sqrt{F+\dot R^2}} - \frac{\ddot R}{\sqrt{1+\dot R^2}} \nonumber\\
&& + \frac{\sqrt{F+ \dot R^2} - \sqrt{1+\dot R^2}}{R} = 0, \label{sangC}
\end{eqnarray}
while
\be
	- 4\pi\sigma = \frac{\sqrt{F+\dot R^2}-\sqrt{1+\dot R^2}}{R} \label{tautau}
\ee
directly describes the evolution of the surface density. For a collapse without change in the rest mass $m=\mathrm{const}$, we have
\be
	r_\sg=2 m\sqrt{1+\dot R^2}-m^2/R.
\ee

The trajectory is obtained by integration of
\be
\dot R=-\sqrt{\left(\frac{r_\sg}{2m}+\frac{m}{2R}\right)^2-1}. \label{clavel}
\ee
Using this result, the equation of motion can be rewritten as
\be
\ddot R=
-\frac{1}{4 R^2}\left (r_\sg+\frac{m^2}{R}\right).
\ee
The integration results in the infinite coordinate time (physical time of a distant observer Bob) and finite proper time of a comoving observer (Alice).

Now, we describe the extensions of this model that include the mass loss by the shell. The exterior geometry is modeled by the standard outgoing Vaidya metric in Sec.~\ref{shell-plus} and by the ingoing Vaidya
metric with decreasing mass in Sec.~\ref{nege}.
\subsection{Evaporating shell and positive energy density}\label{shell-plus}

The outgoing Vaidya metric with $M_u<0$ is an excellent approximation to the  geometry of an evaporating black hole   for $r>3r_\sg$ \cite{bmps:95}. It is
used to model the effects of backreaction in various settings
\cite{kmy:13,ho:16a,pw:99,bmt-1}. 
However, the resulting energy-momentum tensor does not violate the NEC and thus cannot represent an immediate neighborhood of the trapped region that has formed in
 finite time of a distant observer. The presence of evaporation modifies the equation of motion,
\be
{\mathcal{D}(R) }+\frac{F_U}{F\sqrt{F+\dot R^2}}\big(\half-\dot R\dot U\big)=0 ,  \label{eom-u}
\ee
while Eq.~\eqref{tautau} still holds \cite{bmt-1,mnt}.

\begin{figure}[htbp]
	\includegraphics[width=0.45\textwidth]{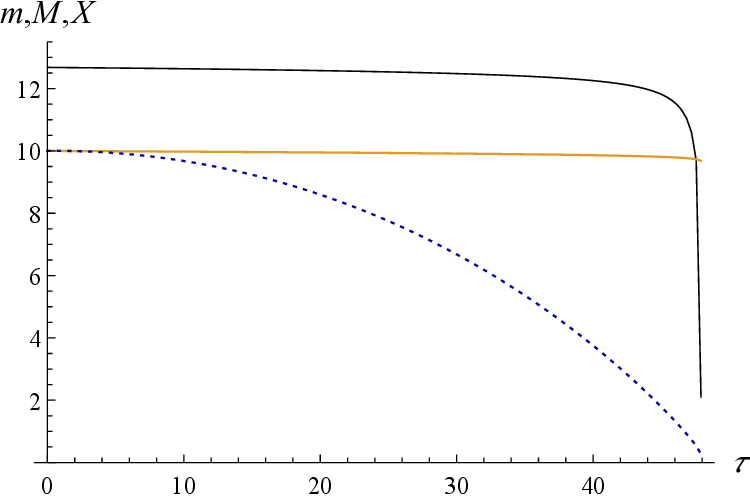}
	\caption{Transition to the null trajectory. The orange line represents the gravitational mass $M(\tau)= r_\sg(\tau)/2$.
 The rest mass $m(\tau)$ is shown as the black line, and the gap $X(\tau)=R(\tau)-r_\sg(\tau)$ as the dotted blue line.
 For simulation purposes, the evaporation was switched on at $\tau=u=0$. The initial conditions are  $R(0)=30$  and $\dot R(0)=0$,
 while $r_0 \vcentcolon= r_\sg(0)=20$, and $\kappa=1$. The evaporation ends at the retarded time $u_*=r_0^3/3\kappa=8000/3$, but the system breaks
 down at approximately $\tau_m=47.912$, indicating a transition to the null trajectory. At the transition, most of the gravitational mass
 is still contained within the shell, $M(\tau_m)/M(0)=0.968$, while the gap is $X(\tau_m)= 0.279$. The closest approach \cite{mnt} is
 determined by $\epsilon_*=0.1$, giving the estimate for the gap at the transition as $X_\infty=0.25$.}
	\label{gap-ur}
\end{figure}

In the limit of large $\dot R$, the asymptotic expression becomes
\be
\ddot R\approx\frac{8 M_U\dot R^4}{RF^2}\approx \frac{16MM_U\dot R^4}{X^2}, \label{ddrap}
\ee
where the second  {relation} in Eq.~\eqref{ddrap} holds for $X\ll r_\sg$. In fact, this accelerates the collapse, as can be seen in Fig. \ref{gap-ur}.

Despite this acceleration, the shell never crosses the ever-shrinking Schwarzschild sphere at $r=r_\sg$, as can be readily deduced by monitoring the previously defined gap (coordinate difference) between the shell and the Schwarzschild radius,
\be
	X \vcentcolon= R-r_\sg.
\ee
Anticipating the transition to a null trajectory, we use a generic parameter $\lambda$ to describe the shell. It can be $u_-$ \cite{cuwy:18} or $R$ itself,
 as is common practice in the analysis of null shells \cite{mnt}.  Using Eq.~\eqref{uder}, we find that close to $r_\sg$
\be
U\approx-\frac{2R_\lambda}{F}\approx - \frac{2R_\lambda r_\sg}{X},
\ee
where the first  relation is exact for null shells and the second is valid for $X \ll r_g$.

Evaluating the derivative of $X$ over $\lambda$, we have
\begin{align}
X_\lambda=R_\lambda -\frac{dr_\sg}{dU}U_\lambda&=-|R_\lambda|\left(1-\left|\frac{dr_\sg}{dU}\right|\frac{2}{F}\right)  \nonumber \\
&\gtrsim -|R_\lambda|\left(1+\frac{8|M_U| M}{X}\right).          \label{avoid}
\end{align}
As a result, the gap decreases only until $X\approx \epsilon_*$ \cite{kmy:13,ho:16a,bmt-1}, where
\be
\epsilon_* \vcentcolon= 2\frac{dr_\sg}{dU}r_\sg=8M|M_U|.
\ee

However, horizon avoidance comes with a price: the shell sheds its rest mass and becomes null in finite proper time.
 This was demonstrated in Ref.\ \cite{cuwy:18} for the outgoing Vaidya metric and in Ref.\ \cite{mnt} for a general metric of the form of Eq.~\eqref{efret}
 with $h(u,r)<+\infty$ and a general evaporation law. While the rest mass {$m=4\pi \sigma R^2$} becomes zero for a finite value of $R>r_\sg>0$, for a macroscopic
  shell, only a negligible fraction of the gravitational mass is lost up to the transition, i.e., $r_\sg\approx r_\sg(0)=\mathrm{const}$. Figure \ref{gap-ur} illustrates
  this process for the Page-like evaporation law
\be
\frac{dr_\sg}{du}=-\frac{\kappa}{r_\sg^2}. \label{evlaw}
\ee

Using the asymptotic form of the equation of motion as given by Eq.~\eqref{ddrap}, we can estimate the transition radius. Since $r_\sg\approx r_0$,
we find $\dot X\approx \dot R$, and thus
\be
\ddot X=-\frac{2\epsilon_*\dot X^4}{X^2}. \ee
The equation for $\tau(X)$ has a simpler form and can be solved by separation of variables if we approximate $\epsilon_*\approx\mathrm{const}$. The first integration gives
\be
\frac{d\tau}{dX} =-\frac{\sqrt{2(-2\epsilon_*+X K_1)}}{\sqrt{X}},
\ee
where $K_1$ is the integration constant. Motion of the shell is affected by evaporation only at distances of the order of $X\ll r_\sg$.
Within the range $r_\sg\gg X\gg \epsilon_*$, where the equation above is already applicable, we set the initial value of $d\tau/dX$
using the classical value of the radial velocity of the shell at the horizon crossing. From Eq.~\eqref{clavel}, we find $\dot R\sim -3/4$ for
a shell initially at rest, giving $K_1=8/9$. At the timelike-to-null transition, $d\tau/dX\to 0$, and therefore it occurs at
\be
X_\infty\approx 2\epsilon_*/K_1=9\epsilon_*/4.  \label{null-t}
\ee

At this point, the model must be supplemented by additional considerations since the unmodified dynamics would inevitably cause the shell to become tachyonic.
There are three main scenarios
that avoid the tachyonic solution: termination of the radiation (i.e., the metric outside of the shell reverts to the Schwarzschild metric in EF coordinates)
\cite{cuwy:18}, modification of the
metric such that the junction conditions for null shells are satisfied from the transition point onward, or preservation of the Vaidya form of the metric and
development of pressure that
 allows to maintain the null trajectory \cite{mnt}.

The first option restores the classical collapse that is completed in finite proper time of Alice and infinite proper time of Bob. In the second case, the
final fate of the shell depends on the specific form of the new metric. The last option leads to horizon avoidance with or without the appearance of a transient naked singularity.

\subsection{Evaporating shell and negative energy density}   \label{nege}
Geometry near the apparent horizon of the contracting trapped region that forms at some finite $t_\mathrm{S}$  is described by the ingoing Vaidya
metric with decreasing $r_\sg(v)$. Using it for the exterior geometry of the shell results in the equation of motion
\be
  \mathcal{D}(R) 	- \frac{F_V}{F \sqrt{\dot{R}^2 + F}} \big(\tfrac{1}{2} + \dot{R} \dot{V}   \big)=0,  \label{eq:EOMshellVRcoord}
\ee
which differs from Eq.~\eqref {eom-u} in a number of ways.

Close to $r_\sg$ (and for nonzero $\dot R$),
\be
\dot V\approx -\frac{1}{2\dot R}+\frac{F}{\dot R^3}\approx -\frac{1}{2\dot R}+\frac{X}{\dot R^3 r_\sg}.
\ee
As a result, the stopping effect of the evaporation is virtually nonexistent.
\begin{figure}[htbp]
	\includegraphics[width=0.45\textwidth]{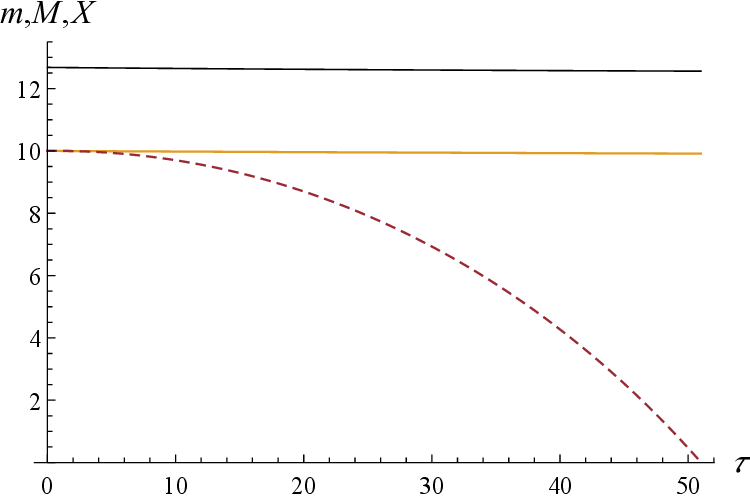}
	\caption{Horizon crossing. The orange line represents  $M(\tau)= r_\sg(\tau)/2$.
The rest mass $m(\tau)$  is shown as the black line, and the gap $X(\tau)=R(\tau)-r_\sg(\tau)$ as the dashed red line.
 For simulation purposes, the evaporation was switched on at $\tau=v=0$. The initial conditions are $R(0)=30$ and $\dot R(0)=0$,
  while  $C_0\equiv C(0)=20$, and $\kappa=1$. At the horizon crossing at $\tau_c=51.010$, the gravitational mass and the rest
   mass are nearly identical to their initial values: $M(\tau_c)/M(0)=0.9913$ and $m(\tau_c)/m(0)=0.9905$, respectively.}
	\label{gap-vr}
\end{figure}

For the shell at rest at $X\ll r_\sg$, assuming again that the evaporation is governed by Eq.~\eqref{page-v},
the radial coordinate acceleration is
\be
\ddot R\approx -\frac{F'}{2}  +\frac{F_V}{F}\approx -\frac{1}{2r_\sg}+\frac{\kappa}{r_\sg^2 X},
\ee
indicating that evaporation prevents the collapse only if
\be
X<\epsilon_*=\frac{2\kappa}{r_\sg}.
\ee

\begin{figure}[htbp]
	\includegraphics[width=0.45\textwidth]{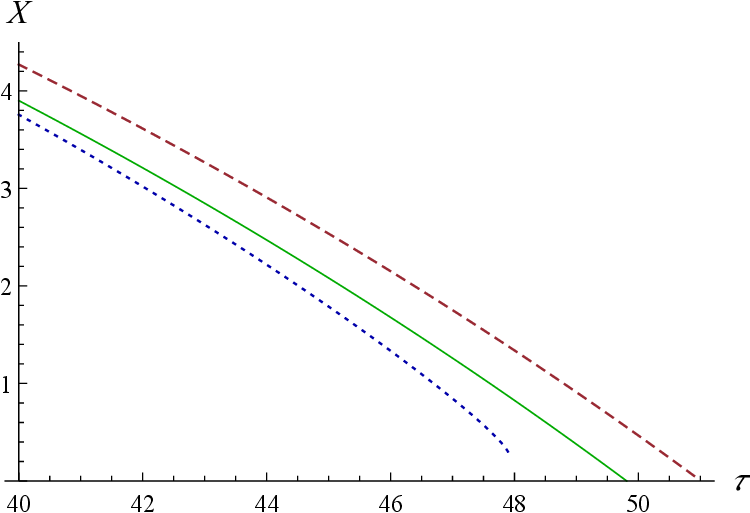}
	\caption{Comparison of the classical trajectory (green line) with the exterior geometry given by the outgoing Vaidya metric with $dr_\sg/du=-\kappa/r_\sg^2$
 (dotted blue line) and the ingoing Vaidya metric with $dr_\sg/dv=-\kappa/r_\sg^2$ (dashed red line) at the later stages of the collapse.
  The initial data is the same as above. In the former case, the collapse is accelerated, and the shell becomes null at $\tau_m$.
  In the latter case, the collapse is slightly delayed.}
 	\label{gap-3}
\end{figure}

It is easy to see that for $\dot R\neq 0$ the (stopping) acceleration term that is proportional to $F_V$ is much smaller than its classical counterpart.
 Figure \ref{gap-vr} illustrates this process for the same value of $\kappa$ and the same initial data as in the previous case. Figure \ref{gap-3}
 illustrates the difference  between classical dynamics and the two  models of a radiating shell.
 Since the influence of evaporation on the dynamics of a macroscopic shell is weak, the shell preserves nearly all of its mass at the horizon crossing.
We can estimate the proper time of the collapse using the classical equation of motion. However, unlike in the classical scenario, the crossing time according to
Bob is finite, as are the propagation time of the last signal that Alice sends before crossing and its redshift.

For illustrative purposes, we approximate the evaporation using a linear law, 
\be
	r_\sg = r_0 - \frac{v}{\zeta} = r_0 - \frac{v \kappa}{r_\sg^2(v=0)} ,
\ee
that allows for explicit analytical results.
Since the shell collapse takes much less time than the evaporation, this is an excellent approximation. 

The equation of an outgoing radial null geodesic
\be
\frac{dv}{dr}=\frac{2}{1-r_\sg(v)/r}
\ee
separates after the change of variables $v=r \tilde v+C_0\zeta$, and its general solution can be written as
\begin{align}
K&=-2\sqrt{\frac{\zeta}{8+\zeta}}\arctanh\left[\sqrt{\frac{\zeta}{8+\zeta}}\left(1+\frac{4 r}{\zeta(r-r_0)+v}\right)\right] \nonumber \\
 &+2 \ln r+\ln\left(1+\frac{\zeta^2 r_0(r-r_0)+\zeta(2r_0-r)v-v^2}{2\zeta r^2}  \right),
\end{align}
where $K$ is the integration constant.

We first use this result to evaluate the redshift that is suffered by the signal sent by Alice at the horizon crossing.
Since the evaporation law is linear, we can adjust parameters such that the crossing happens at $V=0$ at $r_\sg(0)=r_0$.
For the light emitted by Alice at $\Delta\tau$ before the crossing, the constant is given in the leading order by
\be
	K=\pi i-\frac{4\pi i}{\zeta}+2\ln r_0+\frac{\Delta\tau}{r_0}\left(\frac{1}{2|\dot R_0|}+  2|\dot R_0|\right),
\ee
where we also expanded in powers of $\zeta$, using that $\zeta\gg 1$, and $\dot R_0=\dot R(0)$.

\begin{figure}[htbp]
	\includegraphics[width=0.46\textwidth]{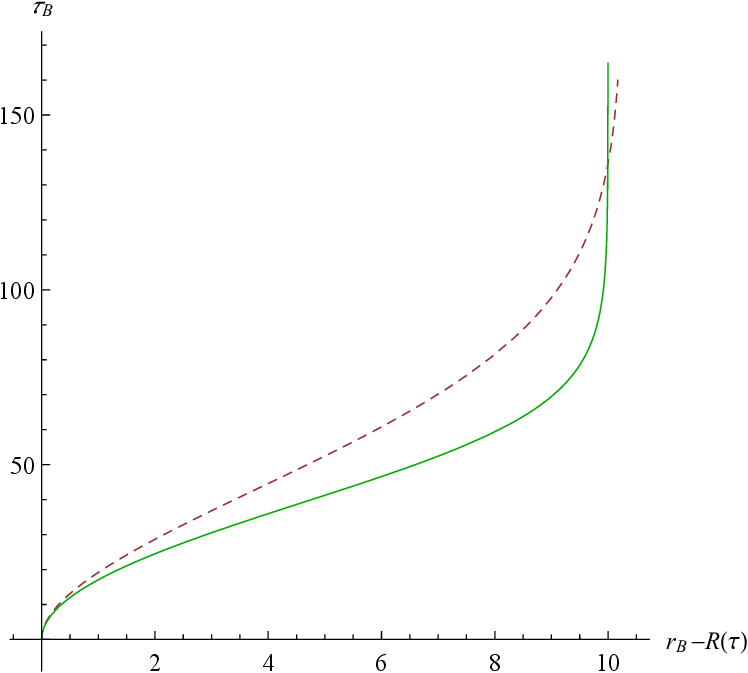}
	\caption{Time of arrival of Alice's signals to Bob. The green line represents the classical case $r_\sg=r_0=\mathrm{const}$.
 The dashed red line represents the case of the linear evaporation law with $\zeta=400$. In this case, the last ray reaches Bob at $\tau_B=160$. }
	\label{time-bob}
\end{figure}

To show that the transmission time and the redshift are finite, we consider a position of Bob that makes the calculations particularly simple.
For this purpose, we locate Bob at the position $r_B$ at which light emitted at $v=0$ reaches Alice at $v_B=\zeta r_0$, i.e., when the shell completely evaporates.
 Consider the beam that arrived at the same location $\Delta v$ earlier. Then, we have (again using $\zeta\gg 1$)
\be
	K=\pi i-\frac{4\pi i}{\zeta}+2\ln r_B -2\ln\zeta +\frac{8\ln\zeta-8-\ln 256}{\zeta}+\frac{\Delta v}{r_B}.
\ee
Comparison of the zeroth-order terms identifies Bob's location,
\be
r_B\approx \half \zeta r_0-(2\ln \zeta -2-\ln 4)r_0+\ldots,
 \ee
while the leading-order term in the redshift is determined by
\be
\frac{dv}{d\tau}=\frac{(1+4\dot R_0^2)r_B}{2 |\dot R_0|r_0}.
\ee
We recall that $|\dot R|\leqslant 3/4$, and since
\be
d\tau_B=\sqrt{f(v,r_B)}dv_B\approx dv_B,
\ee
we see that the redshift is of the order of $\zeta$. If we assume that the initial radius of the shell $R_A$ satisfies $r_0\ll R_A<r_B$, we have $|\dot R_0|=3/4$ and
\be
\frac{dv}{d\tau}=\frac{13}{12}\zeta.
\ee

The time of arrival of the signals that are sent by Alice as the function of her progress is represented in Fig.~\ref{time-bob}. Here, Bob is located at the initial position of Alice,
$r_B=R_A$. His proper time is calculated according to
\be
\tau_B=\int_0^v\sqrt{f(v',r_B})dv'.
\ee
The shell approaches $r_\sg$ in a finite time (both according to Bob and Alice) and with a finite mass. We also observe that the last signal
that is sent by Alice before she crosses the Schwarzschild
sphere reaches Bob in a finite time, while in the classical case, this time diverges as Alice approaches $r_\sg$.

\section{Discussion}\label{summary}

There is no contradiction between the predictions of different models of thin shell collapse, as they are applicable in different situations.
 If the spacetime outside of the shell is described by an
outgoing Vaidya metric with decreasing mass, there is no horizon crossing. This is the expected behavior of a system that satisfies the NEC.
The outgoing Vaidya metric is a good approximation of the geometry around an evaporating black hole for ${r\gtrsim 3 r_\sg}$, while the metric
 approaches the ingoing Vaidya metric with decreasing mass, which is the exact limiting form of the semiclassical geometry \cite{bmmt}, as $r\to r_\sg$.
  A strict interpretation of the QEI estimate \cite{bmmt} indicates that the NEC is violated only up to $x=r-r_\sg\sim a^2\sim \kappa/r_\sg$.
   According to Eq.~\eqref{null-t}, transition to the null trajectory happens at $X_\infty\sim a^2$. Only a more detailed analysis of the geometry
   outside of the collapsing shell, and in particular, finding the details of the transition between the two forms of the Vaidya metric,
    will determine how plausible the scenario of shedding all or most of the shell's rest mass is.
  {Although thin shells can  be used to model radiative processes, either classical or quantum, they cannot accurately describe the final stage of the collapse to a black hole.
  Emission of the entire rest mass may indicate either problems with the thin shell approximation in general or the necessity
  of a more involved specific model for the shell and/or type of radiation. A direct consequence of this result is that the geometry outside of a pressureless null
   shell can never be described by an outgoing Vaidya metric with decreasing mass.

If the geometry outside of the shell is described by an outgoing Vaidya metric with decreasing mass, then the shell
 approaches the Schwarzschild radius in finite time (both according to the comoving Alice and distant Bob),
 losing only an insignificant fraction of its mass. {When it comes to describing the formation of the apparent horizon,
 a potential problem of this idealized model lies in the requirement to alternatingly violate (on both sides of the apparent horizon)
 and satisfy (on both sides of the inner apparent horizon that is the inner boundary of a trapped region propagating to the center) the NEC} \cite{t:19}.
 A thin shell model may be a too extreme idealization to exhibit all of these features, but for $0<R<r_\sg$, a plausible scenario envisages the shell itself
 as the inner apparent horizon, while the trapped region corresponds to $  R<r\leq r_\sg$.

The above results present arguments that thin shell models of collapse and evaporation do not have an independent meaning, but rather
they simply illustrate their underlying assumptions: if the model corresponds to the impossibility of horizon formation at a finite time,
then it will {predict} horizon avoidance. By the same token, if the model uses the metric that is associated with the finite-time appearance of trapped regions,
it {will predict} horizon crossing. This conclusion aligns with the results found in Ref.\ \cite{saini-stojkovic:2015}.

Since the important scales (horizon avoidance, null transition, shell support, and violation of the NEC)  are all of the order of
$X\sim \kappa/M$, we must also consider the possibility that the effect of the transition region between the ingoing and outgoing Vaidya metrics may be such as to prevent the horizon crossing (forcing $\dot X>0$
once some critical value of $X$ has been reached), but without causing the timelike-to-null transition.  This situation will be investigated in future work. The first step to a self-consistent description of gravitational collapse is a complete
 identification of the metric in the neighborhood of the apparent horizon. This was achieved for quasistationary black holes, but our analysis relies on flux calculations on a fixed background.
  The next logical step is to use the pattern provided by the correct self-consistent near-horizon metric to obtain the field modes and derive the renormalized energy-momentum tensor.
  \acknowledgments

We thank Robert Mann for numerous useful discussions, Pisin Chen and Bill Unruh for critical comments, and Joanne Dawson for  helpful suggestions. SM is supported by the iMQRES scheme of Macquarie University.


\end{document}